\documentclass[a4paper]{EslabStyle}
\usepackage{epsfig}

\def\eg{{\it e.g.\,} }

\def\logNHIcm2{\log\left(N_{HI}\,[\mathrm{cm^{-2}}]\right)}

\begin{document}

\title{The Chemical Evolution of the Universe}

\author{A.\,C. Baker\inst{1}\inst{2} \and G.\,P. Mathlin\inst{1} \and D.\,K.
Churches\inst{1} \and M.\,G. Edmunds\inst{1}}

\institute{Department of Physics and Astronomy, Cardiff University, PO
Box 913, Wales, UK CF24 3YB \and a.baker@astro.cf.ac.uk}

\maketitle 

\begin{abstract}

We are encouraged by the improving abundance measurements for quasar
damped absorption line systems to start a new study of galaxy chemical
evolution from high to low redshifts. Our goal is a simple, robust
model based on a synthesis of our best understanding of the relevant
physics, such as cosmology, large-scale structure formation, galaxy
collapse, star formation and chemical evolution. Our initial model
assumes continuous quiescent star formation, and no galaxy mergers. We
use this galaxy chemical evolution model to study the properties high
column density gas out to high redshifts, as revealed by damped
Ly$\alpha$ absorption (DLA). We find that there is a significant
population of {\it dwarf} galaxy DLA systems at moderate redshift ($z
\sim 2.5$, which would not have been accessible to previous imaging
surveys around DLA quasars. Our model also provides a reasonable
prediction of the global star formation history of the Universe as
measured from the UV luminosity density.  The next theoretical step is
clearly to allow galaxy mergers and hence, model the effects of major
bursts of (probably dust-enshrouded) star formation. We are also using
time-delayed element production to probe stellar populations in DLA
galaxies.

\end{abstract}

\section{Introduction}
\label{s:intro}

Some of the most detailed observations which we can make of external
galaxies are spectroscopic, yielding reliable measurements of gas
phase metallicities. Of particular interest are the systems with high
neutral hydrogen column densities which are found {\it via} Ly$\alpha$
absorption lines in the spectra of high redshift quasars: the damped
Ly$\alpha$ absorbers (DLAs). These DLAs may be signatures of normal
galaxies at redshifts $ 0.1 < z < 4.5$, which we can only detect
because of the chance alignment with a background quasar along our
line-of-sight. Because of the high overall gas column density ($20 <
\logNHIcm2 < 21.8 $), it is possible to detect numerous metal line
species and hence infer a range of metal abundances. The database of
DLA abundance data in the literature is growing steadily (see
\cite{M00} and references therein). Meanwhile, major technological
advances are underway, with numerous eight-metre class optical
telescopes and adaptive optics systems being commissioned which will
greatly improve our observations of distant galaxies. The time is ripe
for a new synthesis of our knowledge of the formation and evolution of
galaxies, in order to produce robust theoretical predictions which can
be confronted by these exciting new observations. This contribution
outlines such a project, which is described in much more detail in the
forthcoming paper \cite{M00}.

(Note: All calculations assume an Einstein-de Sitter cosmology
($\Omega=1$) with a present-day Hubble parameter $H_0 = 100 h$
km$^{-1}$ s$^{-1}$ Mpc$^{-1}$ determined by $h=0.7$, unless otherwise
stated.)

\section{A Simple, Robust Model}
\label{s:model}

We do not yet have a complete physical understanding of many of the
mechanisms and processes by which galaxies evolve in their element
abundance makeup.  A particularly important area is the physics of the
formation, evolution and death of stars, and the consequences of
stellar processing for evolution of the gas in galaxy potential wells.
Our cosmological models of the Universe, and of the formation of large
scale structure, and galaxies, are also uncertain.  However, a number
of very simple, yet surprisingly robust `preliminary' models or
parameterisations do exist. Therefore, to build a chemical evolution
model of high redshift galaxies, we have selected the closest to a
`consensus' model which exists in each astrophysical area. We use an
Einstein-de Sitter cosmology (Section~\ref{s:cosmology}), a Schmidt
Law for star formation (Section~\ref{s:starform}), assume constant,
universal stellar yields, and we use the Simple Model of chemical
evolution (Section~\ref{s:chemevol}). We also make a number of
simplifying assumptions at this initial stage, which are being
investigated and relaxed where appropriate as the model is tested and
developed. We assume a universal baryon fraction by mass $f_b = 0.1$,
that the internal structure of galaxies can be neglected, that there
are no gas flows (\eg inflows or outflows, such as accretion,
superwinds or mergers), that all elemental species are `prompt' (no
time delays nor secondary element nucleosynthesis), and in later
stages, that the Holmberg size-luminosity relation for galaxies is
valid (Section~\ref{s:holmberg}). Each of these assumptions is
discussed in detail in \cite{M00}. This is an extremely simple model,
which should however be robust and easy to interpret, because the
input assumptions are clear, and well understood.

\subsection{The basics of the model}
\label{s:basics}

We start by constructing the simplest physically reasonable model of a
galaxy, which is a homogeneous spherically-symmetric cloud of matter,
described by a total mass $M$, and an initial radius $R_0$. The
initial mass of baryons $M_b = f_b M = 0.1 M$. We study model galaxy
clouds with total gravitational masses in the range $10^8 <
M/M_{\odot} < 10^{12}$, and the baryonic component is assumed to be
primordial gas, which can form stars and contribute to the column
density. 

\subsubsection{Cosmology}
\label{s:cosmology}

We initially adopt the simplest cosmological model which is commonly
considered, the Einstein-de Sitter universe. This is the `critical'
Universe, which is exactly closed by the matter density so that
$\Omega=1$.  In line with the best observational evidence, we adopt
Hubble parameter $H_0 = 70 $ km$^{-1}$s$^{-1}$Mpc$^{-1}$. We assume
that structure forms in a `bottom-up' fashion.

\subsubsection{Galaxy formation}
\label{s:galform}

Each model galaxy cloud will `form' when it breaks away from the
Hubble expansion, and starts collapsing to form a gravitationally
bound system, at a redshift which is determined by its fractional
overdensity (relative to the average density of the Universe at that
epoch).  Small length scales will `turn around' first (see \cite{P93}
for extensive discussion). We halt the collapse of each galaxy at $r =
0.1 R_0$, and allow each galaxy to continue to evolve at constant
size. We do not allow galaxies to interact or merge, which is an
important assumption which has significant consequences for the
chemical evolution of each galaxy (Section~\ref{s:chemevol}), and for
the global star formation history of the model Universe (see
Section~\ref{s:madau}). There is as yet no known galaxy initial mass
function, but where necessary, we will simulate one using the the
Holmberg Relation between galaxy size and luminosity
(Section~\ref{s:holmberg}).

\subsubsection{Star formation}
\label{s:starform}

We allow the baryonic gas in the model galaxy clouds to form stars
according to a very simple Schmidt law, dependent only upon total gas
volume density. Kennicutt has demonstrated that such a prescription is
a remarkably good representation of star formation rates under a wide
range of physical conditions in present day galaxies. We adopt the form

\begin{equation}
\label{e:schmidt}
\Gamma = \kappa \rho^{n}_{gas}
\end{equation}
 
where $\Gamma$ is the star formation rate density, and we set the
constant $n=1.5$ (guided by \cite{K98}). We also set the constant
$\kappa = 1.45 M^{-0.5}_{\odot} $pc$^{1.5} $Gyr$^{-1}$ (consistent
with \cite{C99}), and resist the temptation to fine-tune this
parameter, since we have no astrophysically robust prescription for
such variations. We have used present day galaxy gas fraction
measurements to guide us in our choice of $\kappa$, which results in
reasonable values across the whole range of initial galaxy cloud
masses we consider.

We do not impose a specific stellar Initial Mass Function
(IMF). However, we assume a constant, universal yield of metals in our
chemical evolution prescription (Section~\ref{s:chemevol}), which is
consistent with a constant, universal IMF.

As part of our galaxy evolution models, we automatically keep track of
the total mass locked up in long-lived stars and stellar remnants, and
we also calculate the instantaneous mass passing through the current
generation of short-lived stars. This immediately allows us to make a
crude estimate of the relative blue and red luminosities of each model
galaxy cloud at each redshift. A more sophisticated analysis will use
stellar synthesis codes.

\subsubsection{Chemical evolution}
\label{s:chemevol}

We use the Simple Model of chemical evolution (\cite{E90,P97}). We
assume that, in each generation of stars, a fraction $\alpha$ of the
gas which was consumed becomes locked up in long-lived stars and
stellar remnants, and is hence removed from the gas phase and is not
available to subsequent generations of star formation. We assume each
galaxy cloud is a `closed box', so that there are no inflows nor
outflows of any gas.  This is consistent with our assumption that
there are no galaxy interactions nor mergers. We use the instantaneous
recycling and complete mixing approximations, so that the heavy
elements created by each stellar generation are immediately and
uniformly redistributed through the gas cloud.

\subsubsection{Model Galaxy Cloud Evolution}

Taking all these ingredients, we can allow model galaxy clouds to form
and evolve, and track the changes in metallicity and column density as
cosmic time elapses (see \cite{M00}). Therefore, we can now calculate
the key observable physical parameters of DLAs, the absorption
redshift $z_{abs}$, the neutral hydrogen column density $N_{HI}$ and
the metallicity [Z/H]. There is an implicit fourth observable
parameter, the optical depth $\tau$ of the DLA system along our
line-of-sight, which must be sufficiently low to enable us to detect
the background quasar (see Section~\ref{s:dust}).

\section{Comparison with Observations}
\label{s:obs}

There are about 100 DLAs in the literature for which absorption
redshift, HI column density and (in most cases) some measure of metal
abundances are catalogued. In particular, thanks to the concerted
efforts of Pettini and collaborators, there are approximately 40 DLAs
with measurements or limits on the gas phase zinc abundance\footnote{
[Z/H] $= \log \left( n_Z / n_H \right) - \log \left( n_Z / n_H
\right)_{\odot}$ where $n_Z$ is the number of atoms of Z, and $\odot$
indicates Solar values.} [Zn/H].  In principle, we would like to use
iron and oxygen as our primary abundance indicators. However, few
observations of [O/H] have yet been recorded for DLAs. As far as
[Fe/H] is concerned, there are extensive DLA measurements, but it is
likely that iron is significantly depleted into the solid dust phase,
and therefore, the observed gas phase [Fe/H] is very likely an
underestimate of the true metallicity (\cite{P94}). It is though that,
as an iron group element, zinc tracks iron during stellar
nucleosynthesis. It is also known that zinc is significantly less
depleted than iron, and therefore should more accurately probe the gas
phase abundance. This is why there has been a campaign to obtain
[Zn/H] measurements for DLAs, and why we are utilising these data for
comparison with our theoretical chemical evolution predictions. An
important caveat is that the precise stellar origin of zinc remains
unclear, and it may have significant prompt and delayed contributions
(\cite{P97}).

\begin{figure}[ht]
  \begin{center}
	\resizebox{\hsize}{!}{\includegraphics{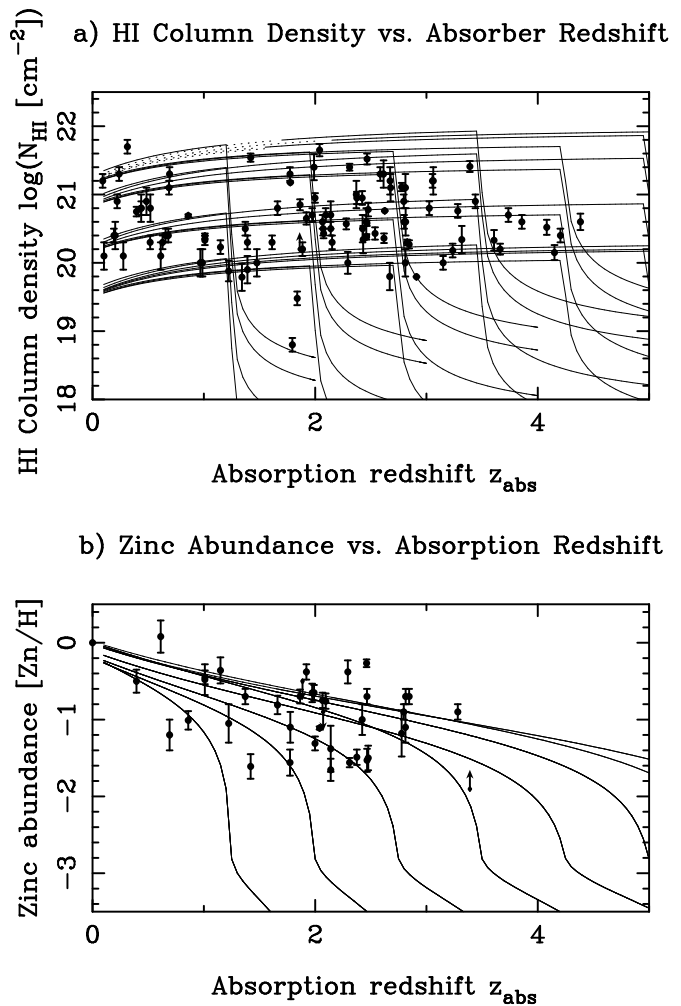}}
  \end{center}	
\caption{\label{f:tracks} A comparison between the properties of
observed damped Ly$\alpha$ systems and the evolution of individual
model galaxy clouds. All the data available in the literature are
shown as data points, and the model evolution tracks are overplotted
(using a dashed line if the cloud is optically thick in dust (see
Section~\ref{s:dust})).  (a) HI column density $N_{HI}\,
\mathrm{cm}^{-2}$ (b) Metallicity measured from zinc ([Zn/H])}
\end{figure}

We can therefore directly compare the column density and metal
evolution of some of our model galaxy clouds with the DLA data, as
shown graphically in Fig.~\ref{f:tracks}. In panel a), the model
galaxy cloud tracks start when the protogalaxy detaches from the
Hubble expansion, and are shown as dashed lines when we estimate that
the cloud will become optically thick (presumably obscuring the
background quasar, and hence becoming undetectable). Each track is for
a cloud of different mass (increasing to higher column density)
forming at a different redshift.  In panel b), we have used the
observed zinc abundance as a probe of the overall metallicity, and
each track represents a different formation redshift (mass is
degenerate in this parameter space, because the chemical evolution
depends only upon density through the Schmidt star formation law).

It is clear that the most simple model produces galaxy cloud evolution
which comfortably explores most of the observed regions of (redshift,
column density, metallicity) parameter space for DLAs.  There are just
a few systems with zinc metallicities which are higher than can be
explained by any of our model galaxy cloud evolution tracks, mostly in
the region $2 < z < 3$.

\subsection{Galaxy gradients in dusty disks}
\label{s:gradients}

The next stage in the development of our theoretical model is to drop
the assumption that galaxy internal structure is negligible. We choose
to do this by allowing our model galaxy clouds to possess gas,
metallicity and dust gradients as have been observed for nearby disk
galaxies. 

\subsubsection{Exponential radial gradients}

We assume purely radial variations, and parameterise the abundance
gradient as $\log Z(r) = a - br $ where Z is the fraction of metals by
mass, $a$ is the central metal fraction, and the slope $b
\sim 0.2 $ dex/scalelength. We calibrate this relationship against our
Galaxy by noting that $ Z_{\odot} \sim 0.017 $ at the Solar circle
$r_{\odot} \sim 8$ kpc. We adopt a universal galaxy stellar scalelength
of $h_* \sim 7$ kpc, typical of giant disk galaxies, and assume that
the gas scalelength is the same as the stellar scalelength,
$h_{HI}=h_*$.

\subsubsection{Estimating dust opacity}
\label{s:dust}

If a DLA is to be detected, there must be a suitable background quasar
in the catalogues, which are often compiled at blue wavelengths. The
quasar must also be bright enough to allow high resolution, high
signal-to-noise spectroscopy. The optical depth of the foreground DLA
galaxy must therefore not be sufficient to redden and obscure the
background quasar beyond the quasar survey and spectroscopic magnitude
limits.

We can easily estimate the maximum dust optical depth of our model
galaxy clouds as a function of redshift and galaxy radius, using the
column density, gas fraction by mass, and radial gas gradient (see
\cite{E98}). We assume that the dust properties are similar to
Galactic dust at all redshifts, and we assume that a fixed fraction
($\eta \sim 0.5$) of metals will condense into the solid phase. The
maximum dust opacity then just a function of the gas column density
and the gas fraction, and so the opacity gradient will track the gas
gradient.

This calculation implies that the {\it maximum} dust opacity through
the disk in the centre of our Galaxy is $\tau_{dust} \sim 6$, which is
a believable firm upper limit, despite uncertainties in the central HI
column density in galaxies. We also predict $\tau_{dust} < 1.3$ at the
Solar circle, which is consistent with observations.

\subsubsection{The properties of DLA galaxies}

We now have a family of homogeneous model galaxy cloud evolutionary
tracks, and a prescription for the radial variation in gas, metal and
dust properties within each galaxy. We therefore have a large set of
possible DLA properties, which are defined by putting a random
line-of-sight through any model galaxy cloud, and reading off the
properties at the radius of intersection. By relaxing the assumption
of cloud homogeneity, the predicted DLA parameters encompass the
region below the fiducial tracks in Fig.~\ref{f:tracks}, reaching
$\log N_{HI} = 20$, and $\Delta\left(\mathrm{[Zn/H]} \right) = 0.6$
respectively.

\begin{figure}[ht]
\begin{tabular}{c}
\resizebox{0.9\hsize}{!}{\includegraphics{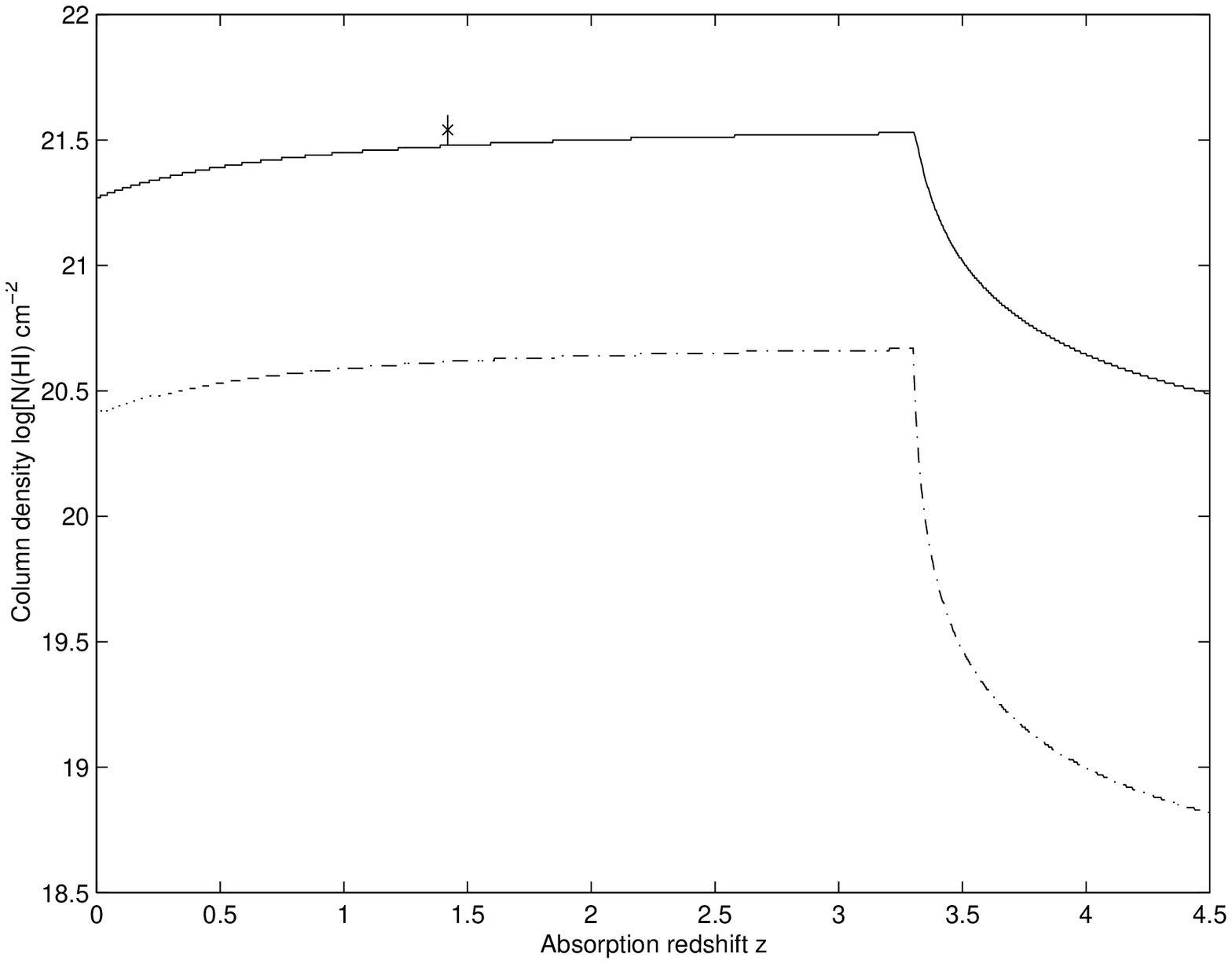}} \\ 
\resizebox{0.9\hsize}{!}{\includegraphics{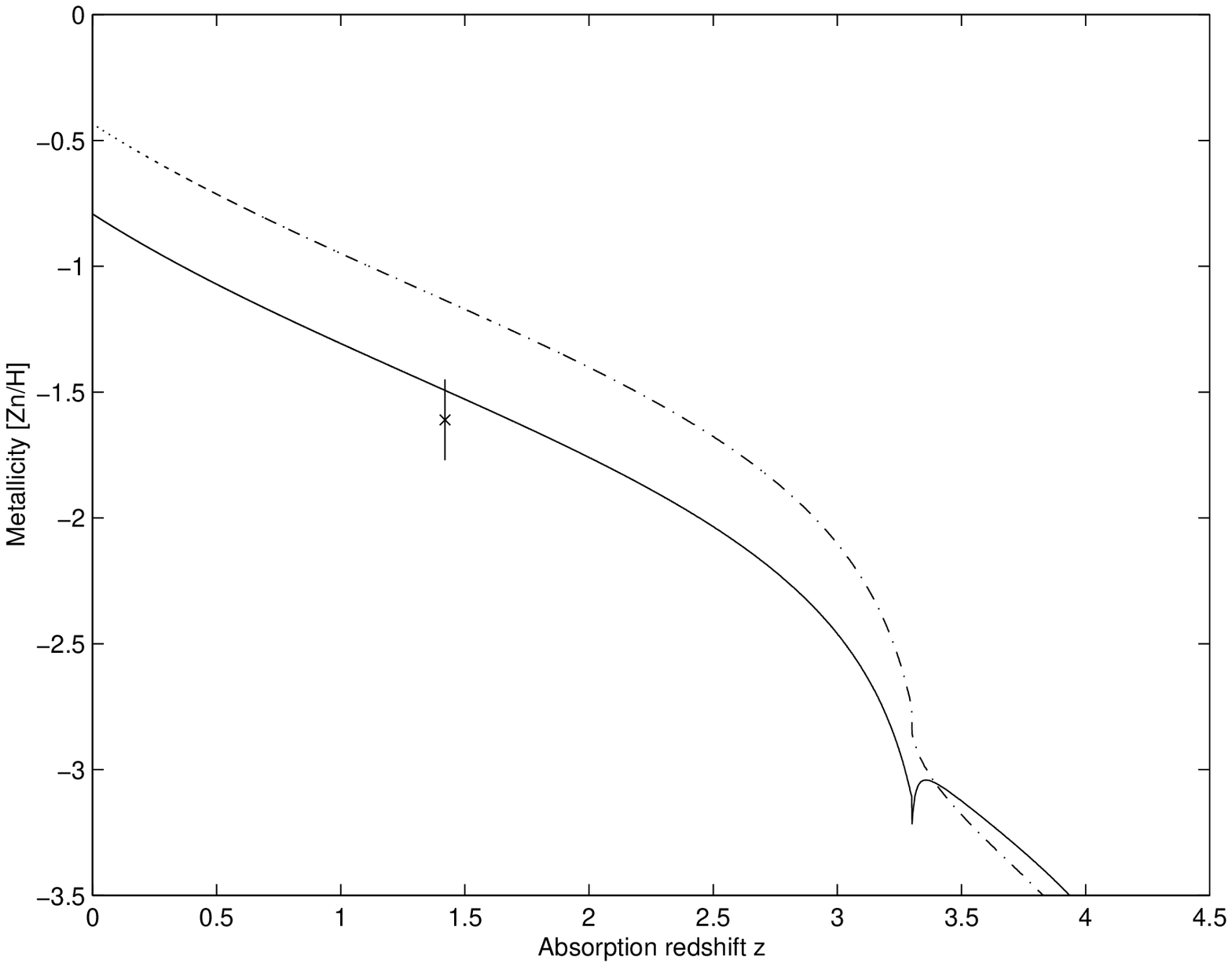}}
\end{tabular}
\caption{\label{f:example} The evolutionary tracks for the best fit
model galaxy cloud ($M_b \sim 10^{10} M_{\odot}$, solid line) for the
absorber seen towards Q$1354+258$ (data points). Also shown are the
evolutionary tracks for a typical dwarf galaxy, which cannot fit this
system ($M_b \sim 10^{7} M_{\odot}$, dot-dash line).}
\end{figure}

In principle, these theoretical sets of parameters might not
correspond to anything which is actually observed.  The predictions
might also be degenerate, with numerous model galaxy clouds
reproducing one column density and metallicity pair at a given
redshift. But in fact, for almost every known DLA, we have been able
to find a unique model galaxy cloud which has the appropriate column
density and metallicity at the observed redshift, within the
observational error bars. An example is shown in Fig.~\ref{f:example},
for a DLA observed in the spectrum of the quasar Q$1354+258$.

We have therefore reached an important milestone in our study. We can
now take the observed properties of a damped Ly$\alpha$ absorption
system, and use our model to make some definite predictions about the
properties of the underlying galaxy. In particular, for Q$1354+258$,
we predict relatively massive galaxy, with a baryonic mass $M_b \sim
10^{10} M_{\odot}$ (and so, a total gravitational mass $M \sim 10^{11}
M_{\odot}$) which formed at a redshift $z_{turn} \sim 4.5$. As noted
in Section~\ref{s:starform}, we can also roughly calculate the
luminosity of the young and old stellar populations.  For Q$1354+258$,
we predict $B \sim 26$ and $K \sim 20$. We also predict that the
nucleus of the DLA galaxy lies about $5$kpc from our line-of-sight to
the quasar.

We now know where to look on the sky, and in blue and red apparent
magnitude space, to try and directly detect the DLA galaxy
itself. These are very interesting predictions from an observational
point-of-view. Furthermore, some interesting patterns start to emerge
as we extend the modelling to all the individual DLAs for which [Zn/H]
measurements exist. Approximately half of all the underlying DLA
galaxies are far less massive than that implicated in the Q$1354+258$
system. We therefore predict that approximately half of DLAs in the
redshift range $ 1 < z < 3$ are due to absorption occuring in {\it
dwarf} galaxies, in stark contrast to the `standard model' of DLA
galaxies as the progenitors of present day giant disk galaxies. This
in turn implies that although the {\it majority} of DLA galaxies are
going to be exceedingly faint at {\it optical} wavelengths, a
significant fraction should be detectable in the {\it near-IR}, above
$K \sim 20.5$. Our model allows us to clearly predict which DLA
galaxies are mostly likely to be accessible to currently observational
technology. We are therefore in a position to use our model to
construct a well-motivated experimental observational programme, to
make high spatial resolution near-IR imaging observations of carefully
selected DLA quasars. We hope to be able to report the results of such
a programme soon.

\subsection{The distribution of DLAs in parameter space}

What is the probability that a random line of sight will pass through
a damped Ly$\alpha$ galaxy with a given column density at a given
redshift? We can make a rough calculation, since we know the
cross-sectional area and column density of our model galaxy clouds,
given that we have a cosmological model and a galaxy initial mass
function. Since the galaxy IMF is not yet known, some other prediction
of a galaxy mass-formation relation, or equivalently, a galaxy
mass-radius relation, must be found.

\subsubsection{The Holmberg Relation}
\label{s:holmberg}

There does exist a well-known linear correlation between galaxy
absolute magnitude and the logarithm of galaxy radius, as originally
described by Holmberg (1975). We have converted this into a galaxy
mass-radius relationship by assuming a constant, universal
mass-to-light ratio. This means that all galaxies of a given mass have
the same present-day radius, and hence, the same turn-around radius
$R_0$, the same turn-around density, and the same turn-around or
formation redshift. We can now construct a statistically meaningful
Universe where the galaxies of the lowest mass turn around at the
highest redshift. We constrain our Universe to reproduce the Schechter
form of the observed present-day overall galaxy luminosity function
(\cite{B98}). Therefore, as cosmic time advances, we `march down' the
Schechter function, forming fewer and fewer galaxies of larger and
larger mass as we approach $z=0$.

A serious caveat is that there are strong indications that the
Holmberg relationship may be an artifact of a combination of
observational selection effects. There are reasons to believe that
compact, luminous galaxies are missed in surveys because they are
mistaken for stars, and that diffuse, faint galaxies (the famous Low
Surface Brightness galaxies) are also missed against the sky
background.  An important development of our theoretical framework
will be to discard the Holmberg Relation and find a better
prescription for the galaxy birth function.

\subsubsection{Random lines of sight}

\begin{figure}[ht]
  \begin{center}
	\resizebox{\hsize}{!}{\includegraphics{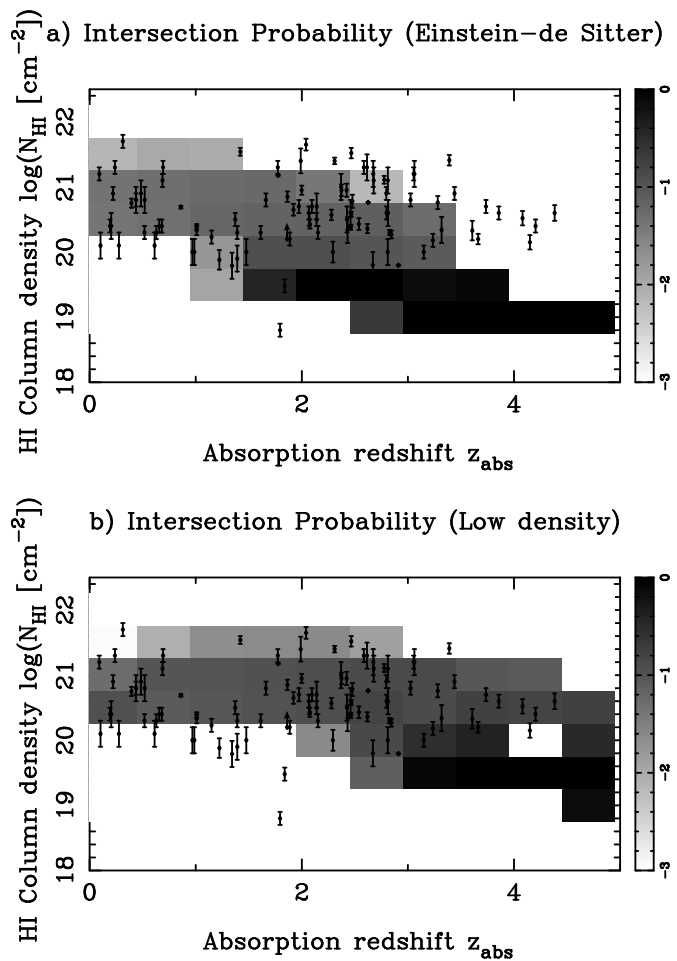}}
  \end{center}	
\caption{\label{f:probs} The data are given by the points (see
Fig.~\ref{f:tracks}). A rough indication of the (logarithm of the)
probability that a random line of sight will intersect a model damped
Ly$\alpha$ galaxy cloud of a given column density at a given redshift
shown by the greyscale. Two different cosmologies (time-redshift
relations) are considered: a) Einstein-de Sitter ($\Omega=1$,
$t_0\sim10$ Gyr) b) Low density ($\Omega=0.3$, $t_0\sim17$ Gyr). Note
that `white' indicates `undefined' in both panels. }
\end{figure}

We have chosen to make one calculation of the probable DLA properties
along a random line of sight in our standard Einstein-de Sitter
cosmology. We have also repeated the exercise for a low density
Universe ($\Omega=0.3$). The results are shown in Fig.~\ref{f:probs},
as a grey scale of the logarithm of the probability that a random line
of sight will intersect at that point in column density-redshift
parameter space. (Note that we cannot calculate the probability across
the entire graph - the regions which are white are undefined because
we do not have model galaxy clouds which track through those parameter
regions.)

It is clear from panel a) that we have extreme difficultly in
explaining the existence of high column density systems at high
redshift in an Einstein-de Sitter Universe with our current model. In
the low density Universe (panel b), we seem to broadly be able to
understand the spread of the observed DLA population. It is tempting
to use this as evidence for or against particular cosmological models.
However, it is vital to recall that in general, it is impossible to
separate cosmological evolutionary effects from the characteristics of
an evolving population of astronomical objects. This can only be done
on the basis of a firm independent understanding of the physics and
evolution of the astronomical objects in question. This trap for the
unwary has resulted in overly confident cosmological pronouncements in
the past.

There are also important subtle observational selection effects
operating in the selection of DLA galaxies. Technological and physical
limitations mean that spectroscopic measurements are best for systems
around $z \sim 1$ (space-based UV data \eg HST) and $2 < z < 3$
(ground-based optical data). Meanwhile, although damping wings start
to appear in the Ly$\alpha$ line at column densities $\log N_{HI} \sim
19$, the effect is very difficult to reliably detect below $\log
N_{HI} \sim 20$.  Also, as alluded to above (Section~\ref{s:dust}),
concerns about the foreground DLA obscuring any background quasar
become significant at the higher column densities. Therefore, we can
only hope for a relatively uniform {\it observational} selection
function in two small `boxes' in column density-redshift space.

There are thousands of quasars in the catalogues, of which fewer than
1\% are currently known to have DLA systems in their spectra. It is
difficult to calculate reliable statistics for the complete population
of known quasars. The shape of our predicted probability distribution
around $z \sim 2.5 $ appears to be broadly correct, rolling off
towards higher column densities, and fairly flat towards the
observational cutoff at $\log N_{HI} = 20 $. However,
we appear to be predicting that 10\% of quasars should have a moderate
redshift DLA system ($ z \sim 2.5 $), which clearly contradicts the
observations.

\subsection{The star formation history of the Universe}
\label{s:madau}

\begin{figure}[ht]
  \begin{center}
	\resizebox{\hsize}{!}{\rotatebox{-90}{\includegraphics{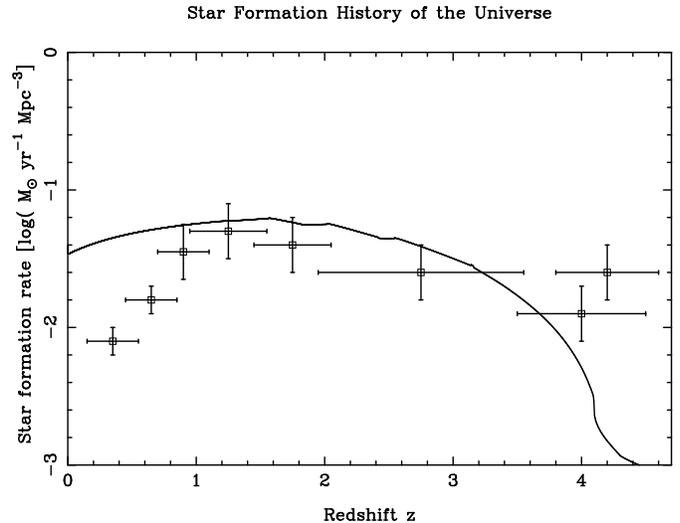}}}
  \end{center}	
\caption{\label{f:madau} The star formation history of the
Universe. The data points are taken from Steidel (1999), from UV
luminosity density observations, with the rough predictions from our
model galaxy clouds overplotted as a (smoothed) curve. This curve has
{\bf not} been fitted to the data.}
\end{figure}

There has been much recent excitement about `measurements' of the
evolution of the star formation rate by mass per co-moving volume with
cosmic time. Once we have specified the birth function of galaxies,
and the star formation law in our model, we too can calculate this
quantity. The results are shown in Fig.~\ref{f:madau}. We have {\it
not} attempted to make a fit to the observational values for the star
formation rate evolution. The rough predictions shown are a
consequence of the parameters which we have chosen for the Holmberg
galaxy size-luminosity relation and the Schmidt star formation law,
subject to the general constraints that present-day galaxies should
exist, spanning the range of mass from dwarf to giant, and should
still contain some hydrogen gas (rather than turning into `ball
bearings' with only metals). 

The comparisons between our rough model predictions and the star
formation rate evolution predicted from measurements of the UV
luminosity density evolution of the Universe are very
thought-provoking. Remember that we have not allowed our galaxies to
merge, and therefore, there is no interaction or merger-induced
starburst activity included in our models. Merger-induced bursts of
star formation are closely linked with highly dust-embedded,
infrared-luminous galaxies (\cite{C96}), which have extremely low UV
luminosities. Much controversy has surrounded the contribution of such
dust-embedded star formation to the overall budget for the conversion
of gas to stars, which must largely be missing from UV based
calculations such as those of \cite{S99}. Since this star formation
mode is not yet included in our models, our rough predictions should
be quite directly comparable to the UV star formation rate
measurements. It is clear from Fig.~\ref{f:madau} that our model
galaxy clouds are forming stars at approximately the same rate as can
be observed in the UV in the real Universe, and during approximately
the same period in cosmic time. Just as clearly, our rough prediction
for the current UV luminosity from star formation is significantly too
high. We could fine-tune this by tuning $\kappa$ in the Schmidt law,
but this would over-predict the star formation rate at higher
redshifts. We could tweak the Holmberg relation to compensate for that
effect, but this will not be very instructive (see
Section~\ref{s:holmberg}).

\section{Conclusions and Further Work}
\label{s:conclusions}

The extremely simple galaxy chemical evolution model which we have
described in this contribution has been remarkably successful and
informative when applied to the study of galaxies detected as damped
Ly$\alpha$ absorption systems in the spectra of quasars.  We have
gained sufficient insight to make definite predictions about the
nature of the DLA galaxies. Our models suggest that, contrary to some
expectations, perhaps half the DLA systems at moderate redshifts ($z
\sim 2.5 $) arise in {\it dwarf} galaxies. These dwarf DLA galaxies
will be faint systems at low impact parameters ($5-10$ kpc) to our
line-of-sight to the quasar. This prediction makes sense of the
relative lack of success of previous imaging searches for
moderate-redshift DLA galaxies (\eg \cite{AEO96}). In our current
theoretical framework, we would expect that at high redshifts ($z>3$),
the population of galaxies selected as DLA absorption systems would be
dominated by dwarf galaxies, since giant galaxies are yet to form. At
low redshifts ($z<1$), the giant galaxies would dominate the DLA
population.

However, at this stage, we have pushed this first model to its very
limits, and now need to start exploring the consequences of relaxing
some of the more stringent assumptions. In particular, the next
theoretical iteration is already underway, to allow mergers between
model galaxy clouds, to find a robust prescription for the galaxy
IMF,and to consider the chemical evolution of delayed element products
of stellar nucleosynthesis. These extra theoretical elements will
enable us to include the effects of merger-induced (probably
dust-enshrouded) star formation in our chemical evolution models, to
make much firmer statistical predictions about the DLA galaxy
population, and to study potential dust, ionisation and stellar
population variations in DLA galaxy evolution. The results will be
presented in \cite{B00}.


\begin{thebibliography}{}

\bibitem[\protect\astroncite{Arag\'{o}n-Salamanca, Ellis \&
O'Brien}{1996}]{AEO96} Arag\'{o}n-Salamanca, A, Ellis, R S, O'Brien, K
S 1996 MNRAS 281 945

\bibitem[\protect\astroncite{Baker et~al.}{2000}]{B00}
Baker, A C, Mathlin, G P, Davey, J, Churches, D K, Edmunds,
M G 2000 MNRAS (in preparation)

\bibitem[\protect\astroncite{Binney \& Merrifield}{1998}]{B98}
Binney, J, Merrifield, M Galactic Astronomy, Princeton University Press

\bibitem[\protect\astroncite{Clements et~al.}{1996}]{C96}
Clements, D L, Sutherland, W J, McMahon, R G, Saunders, W 1996 MNRAS 279 477

\bibitem[\protect\astroncite{Churches}{1999}]{C99} Churches, D K 1999
Numerical Simulations of the Chemical Evolution of Galaxies PhD
Thesis, Cardiff

\bibitem[\protect\astroncite{Eales \& Edmunds}{1998}]{E98}
Eales, S A, Edmunds, M G 1998 MNRAS 299 L29

\bibitem[\protect\astroncite{Edmunds}{1990}]{E90}
Edmunds, M G 1990 MNRAS 246 678

\bibitem[\protect\astroncite{Holmberg}{1975}]{H75}
Holmberg, E 1975 in Galaxies and the Universe, Sandage A, Sandage M,
\& Kristian, J (eds) p154

\bibitem[\protect\astroncite{Kennicutt}{1998}]{K98}
Kennicutt, R C 1998 ApJ 498 541 

\bibitem[\protect\astroncite{Mathlin et~al.}{2000}]{M00}
Mathlin, G P, Baker, A C, Churches, D K, Edmunds, M G 2000 MNRAS (in
preparation)

\bibitem[\protect\astroncite{Pagel}{1997}]{P97}
Pagel, B 1997 Nucleosynthesis and Chemical Evolution of Galaxies,
Cambridge University Press

\bibitem[\protect\astroncite{Peebles}{1993}]{P93}
Peebles, P J E 1993 Principles of Physical Cosmology, Princeton
University Press

\bibitem[\protect\astroncite{Pettini et~al.}{1994}]{P94}
Pettini, M, Smith, L, Hunstead, R W, King, D L 1994 ApJ 426 79

\bibitem[\protect\astroncite{Steidel et~al.}{1999}]{S99}
Steidel, C C, Adelberger, K L, Giavalisco, M, Dickinson, M, Pettini, M
1999 ApJ 519 1

\end{thebibliography}
\end{document}